\documentclass[aps,twocolumn,showpacs,superscriptaddress,groupedaddress]{revtex4}

\usepackage[all, warning]{onlyamsmath}
\usepackage{amssymb, amsmath}
\usepackage{graphicx}
\usepackage{dcolumn}
\usepackage{bm}
\usepackage[figuresright]{rotating}
\usepackage{fancyhdr}
\usepackage{enumitem}
\usepackage{siunitx}
\usepackage{indentfirst}
\usepackage{xspace}
\usepackage{color}
\usepackage{epsfig}
\usepackage{grffile}
\usepackage[T1]{fontenc}
\usepackage[utf8]{inputenc}
\usepackage{color}

\newcommand{\etal}{\textsl{et al}.}
\newcommand{\pizero}{\pi^{0}}
\newcommand{\pt}{p_\text{T}}
\newcommand{\sqs}{\sqrt{s}}
\newcommand{\snn}{\sqrt{s_{NN}}}

\newcommand{\ybeam}{y_\text{beam}}
\newcommand{\ylab}{y_\text{lab}}
\newcommand{\ycms}{y_\text{cms}}

\newcommand{\nmf}{R_\textrm{pPb}}
\newcommand{\mev}[1]{\SI{#1}{\mega\electronvolt}}
\newcommand{\gev}[1]{\SI{#1}{\giga\electronvolt}}
\newcommand{\tev}[1]{\SI{#1}{\tera\electronvolt}}

\begin{document}

\title{Transverse-momentum distribution and nuclear modification factor for
neutral pions in the forward-rapidity region in proton-lead collisions at $\snn
= \tev{5.02}$}
\newcommand{\firenzeinfn}{\affiliation{INFN Section of Florence, Italy}}
\newcommand{\firenzeuniv}{\affiliation{University of Florence, Italy}}
\newcommand{\firenzecnr}{\affiliation{IFAC-CNR, Italy}}
\newcommand{\nagoya}{\affiliation{Solar-Terrestrial Environment Laboratory, Nagoya University, Nagoya, Japan}}
\newcommand{\nagoyaphys}{\affiliation{Graduate school of Science, Nagoya University, Japan}} 
\newcommand{\kmi}{\affiliation{Kobayashi-Maskawa Institute for the Origin of Particles and the Universe, Nagoya University, Nagoya, Japan}}
\newcommand{\polyteque}{\affiliation{Ecole-Polytechnique, Palaiseau, France}}
\newcommand{\waseda}{\affiliation{RISE, Waseda University, Japan}}
\newcommand{\cern}{\affiliation{CERN, Switzerland}}
\newcommand{\kanagawa}{\affiliation{Kanagawa University, Japan}}
\newcommand{\cataniainfn}{\affiliation{INFN Section of Catania, Italy}}
\newcommand{\cataniauniv}{\affiliation{University of Catania, Italy}}
\newcommand{\lbn}{\affiliation{LBNL, Berkeley, California, USA}}

\author{O.~Adriani}
\firenzeinfn
\firenzeuniv

\author{E.~Berti}
\firenzeinfn
\firenzeuniv

\author{L.~Bonechi}
\firenzeinfn

\author{M.~Bongi}
\firenzeinfn
\firenzeuniv

\author{G.~Castellini}
\firenzeinfn
\firenzecnr

\author{R.~D'Alessandro}
\firenzeinfn
\firenzeuniv

\author{M.~Del~Prete}
\firenzeinfn
\firenzeuniv

\author{M.~Haguenauer}
\polyteque

\author{Y.~Itow}
\nagoya
\kmi

\author{K.~Kasahara}
\waseda

\author{K.~Kawade}
\nagoya

\author{Y.~Makino}
\nagoya

\author{K.~Masuda}
\nagoya

\author{E.~Matsubayashi}
\nagoya

\author{H.~Menjo}
\nagoyaphys

\author{G.~Mitsuka}
\firenzeuniv
\nagoya

\author{Y.~Muraki}
\nagoya

\author{P.~Papini}
\firenzeinfn

\author{A.-L.~Perrot}
\cern

\author{D.~Pfeiffer}
\cern

\author{S.~Ricciarini}
\firenzeinfn
\firenzecnr

\author{T.~Sako}
\nagoya
\kmi

\author{N.~Sakurai}
\kmi

\author{T.~Suzuki}
\waseda

\author{T.~Tamura}
\kanagawa

\author{A.~Tiberio}
\firenzeinfn
\firenzeuniv

\author{S.~Torii}
\waseda

\author{A.~Tricomi}
\cataniainfn
\cataniauniv

\author{W.~C.~Turner}
\lbn

\collaboration{The LHCf Collaboration}
\noaffiliation

\date{\today}

\begin{abstract}
The transverse momentum ($\pt$) distribution for inclusive neutral pions in the
very forward rapidity region has been measured, with the Large Hadron Collider
forward detector (LHCf), in proton--lead collisions at nucleon-nucleon
center-of-mass energies of $\snn = \tev{5.02}$ at the LHC. The $\pt$ spectra
obtained in the rapidity range $-11.0 < \ylab < -8.9$ and $0 < \pt < \gev{0.6}$
 (in the detector reference frame) show a strong suppression of the production
of neutral pions after taking into account ultra-peripheral collisions. This
leads to a nuclear modification factor value, relative to the interpolated $\pt$
spectra in proton-proton collisions at $\sqs = \tev{5.02}$, of about 0.1--0.4.
This value is compared with the predictions of several hadronic interaction
Monte Carlo simulations.
\end{abstract}


\maketitle

%
%
\section{Introduction}\label{sec:introduction}

Measurements of particle production in hadronic interactions at high energies
play an unique role in the study of strong interactions described by Quantum
Chromodynamics (QCD). For example, as first discovered in measurements at
HERA~\cite{H1, ZEUS}, it is still not well understood how a parton (dominantly
gluon) density increases or even saturates when the momentum fraction that the
parton itself carries is small (denoted as Bjorken-$x$). This even though the
understanding of the behaviour of hadron constituents (partons) has been
improving both theoretically and experimentally in the past few decades.

Such kind of phenomena at small Bjorken-$x$ are known to be more visible in
events at large rapidities. Since at large rapidities partons in projectile and
target hadrons generally have large and small momentum fractions respectively,
Bjorken-$x$ in the target should be smaller than in mid rapidity events.
Moreover, in case of nuclear target interactions, the parton density in the
target nucleus is expected to be larger by $\sim A^{1/3}$. In these interactions
partons in the projectile hadron would lose their energy while traveling in the
dense QCD matter. Particle production mechanisms change accordingly when
compared to those in nucleon-nucleon interactions.

These phenomena have been observed by several experiments at the Super Proton
Synchrotron (SPS, CERN), at the Relativistic Heavy Ion Collider (RHIC, BNL), and
at the Large Hadron Collider (LHC, CERN). The BRAHMS and STAR experiments at
RHIC showed the modification of particle production spectra at forward
rapidity~\cite{BRAHMS, STAR} by comparing the $\pt$ spectra in deuterium--gold
($d$--Au) collisions at nucleon-nucleon center-of-mass energies $\snn =
\gev{200}$ with those in proton--proton ($p$--$p$) collisions at center-of-mass
energies $\sqs = \gev{200}$. Especially, a comparison of the experimental
results between different rapidity regions by BRAHMS (pseudorapidity $\eta =
2.2$ and 3.2) and STAR ($\eta = 4.0$) indicates that particle production is
strongly suppressed with increasing pseudorapidity.
Also at LHC the same suppression of hadron production has been found by the
ALICE and LHCb experiments, both at mid and forward rapidities~\cite{ALICE,
LHCb}.

Thus one could ask, in what way does particle production take place within an
extremely dense QCD matter in \textit{very forward} rapidity regions ?
There are models that actually try to make quantitative predictions in these
regions. For example under the black body assumption the meson production is
found to be strongly suppressed as a result of limiting fragmentation with a
broadened $\pt$ distribution~\cite{Blackbody}.
A suppression of particle production is also predicted using the color glass
condensate formalism because of the gluon saturation
dynamics~\cite{JalilianMarian, Albacete}.
Similarly hadronic interaction Monte Carlo (MC) simulations covering soft-QCD
show a modification of the $\pt$ distributions.
Experimental data that confirm the theoretical and phenomenological predictions
and possibly constrain the remaining degrees of freedom in such models would
thus be very welcome.

Understanding of particle production in very forward rapidity regions in nuclear
target interactions is also of importance for ultrahigh-energy cosmic ray
interactions, where parton density is expected to be much higher than at LHC
energies due to the dependance of Bjorken-$x \propto 1/\sqs$.
In high energy cosmic-ray-observation, energy and chemical composition of
primary cosmic rays are measured by analysing the cascade showers produced by
the cosmic rays interacting with the nuclei in the earth
atmosphere~\cite{UHECR}. Secondary particles produced in the atmospheric
interaction are, of course, identical to the forward emitted particles from the
hadronic interactions at equivalent collision energy.
In fact current modeling of particle production in nuclear interactions is
limited by the available energy at the accelerators and is the cause of large
systematic uncertainties in high energy cosmic ray physics.

The Large Hadron Collider forward (LHCf) experiment~\cite{LHCfTDR} is designed
to measure the hadronic production cross sections of neutral particles emitted
in very forward angles in $p$--$p$ and proton-lead ($p$--Pb) collisions at the
LHC, including zero degree.
The LHCf detectors (see Sec.~\ref{sec:detector}) cover a pseudorapidity range
larger than 8.4 and are capable of precise measurements of the forward
high-energy inclusive-particle-production cross sections of photons, neutrons,
and other neutral mesons and baryons. Therefore the LHCf experiment provides a
unique opportunity to investigate the effects of high parton density which is
the case in the small Bjorken-$x$ region and in $p$--Pb collisions at high
energies.

In the analysis presented in this paper, the focus is placed on the neutral
pions ($\pizero$s) emitted into the direction of the proton beam (proton remnant
side), the most sensitive probe to the details of the $p$--Pb interactions. From
the LHCf measurements, the inclusive production rate and the nuclear
modification factor for $\pizero$s in the rapidity range of $-11.0 < \ylab <
-8.9$ in the detector reference frame are then derived as a function of the
$\pizero$ transverse momentum.

The paper is organised as follows. Sec.~\ref{sec:detector} gives a brief
description of the LHCf detectors. Section~\ref{sec:data} and
Sec.~\ref{sec:mc_simulation} summarize the data taking conditions and the MC
simulation methodology, respectively. In Sec.~\ref{sec:framework} the analysis
framework is described, while the factors that contribute to the systematic
uncertainties are explained in Sec.~\ref{sec:systerror}. Finally the analysis
results are presented and discussed in Sec.~\ref{sec:result}. Concluding remarks
are found in Sec.~\ref{sec:conclusions}.

%
%
\section{The LHCf detector}\label{sec:detector}

Two independent detectors called the LHCf Arm1 and Arm2 were assembled
originally to study $p$--$p$ collisions at the LHC~\cite{LHCfJINST,
LHCfsilicon}. In $p$--Pb collisions at $\snn = \tev{5.02}$, only the LHCf Arm2
detector was used to measure the secondary particles emitted into the proton
remnant side.
Hereafter the LHCf Arm2 detector is denoted as the LHCf detector for brevity.
The LHCf detector has two sampling and imaging calorimeters composed of 44
radiation lengths of tungsten and 16 sampling layers of \SI{3}{\milli\meter}
thick plastic scintillators. The transverse sizes of the calorimeters are
25$\times$\SI{25}{\milli\meter}$^2$ and 32$\times$\SI{32}{\milli\meter}$^2$.
Four X-Y layers of silicon microstrip sensors are interleaved with the layers of
tungsten and scintillator in order to provide the transverse profiles of the
showers. Readout pitches of the silicon microstrip sensors are
\SI{0.16}{\milli\meter}~\cite{LHCfsilicon}.

The LHCf detector was installed in the instrumentation slot of the target
neutral absorber (TAN)~\cite{TAN} located \SI{140}{\meter} in the direction of
the ALICE interaction point (IP2) from the ATLAS interaction point (IP1) and at
a zero-degree collision angle. The trajectories of charged particles produced at
IP1 and directed towards the TAN are bent by the inner beam separation dipole
magnet D1 before reaching the TAN itself. Consequently, only neutral particles
produced at IP1 enter the LHCf detector.
The vertical position of the LHCf detector in the TAN is manipulated so that the
LHCf detector covers the pseudorapidity range from 8.4 to infinity for a beam
crossing angle of \SI{145}{\micro\radian}, especially the smaller calorimeter
covers the zero-degree collision angle. After the operations in $p$--Pb
collisions, the LHCf detector was uninstalled from the instrumentation slot of
the TAN on April, 2013.

More details on the scientific goals of the experiment are given in
Ref.~\cite{LHCfTDR}. The construction of the LHCf detectors (Arm1 and Arm2) is
reported in Refs.~\cite{LHCfJINST, LHCfsilicon} and the performance of the
detectors has been studied in the previous reports~\cite{SPS2007, LHCfIJMPA}.

%
%
\section{Experimental data taking conditions}\label{sec:data}

The experimental data used for the analysis in this paper were obtained in
$p$--Pb collisions at $\snn = \tev{5.02}$ with a \SI{145}{\micro\radian} beam
crossing angle. The beam energies were \tev{4} for protons and \tev{1.58} per
nucleon for Pb nuclei. Since the beam energies were asymmetric the
nucleon-nucleon center-of-mass in $p$--Pb collisions shifted to rapidity =
$-0.465$, with the proton beam traveling at $\theta = \pi$ and the Pb beam at
$\theta = 0$.

Data used in this analysis were taken in three different runs. The first (LHC
Fill 3478) was taken on January 21, 2013 from 02:14 to 03:53. The second and
third runs (LHC Fill 3481) were taken on January 21, 2013 from 21:03 to 23:36
and January 22, 2013 from 03:47 to 04:48, respectively. The integrated
luminosity of the data was \SI{0.63}{\nano\barn^{-1}} after correcting for the
live time of data acquisition systems~\cite{CERNLumi}.
The average live time percentages for LHC Fill 3478 and 3481 were
\SI{12.1}{\percent} and \SI{6.3}{\percent} respectively, the smaller live time
percentage in Fill 3481 relative to Fill 3478 being due to a difference in the
instantaneous luminosities.
These three runs were taken with the same data acquisition configuration. In all
runs the trigger scheme was essentially identical to that used in $p$--$p$
collisions at $\sqs = \tev{7}$. The trigger efficiency was greater than
$\SI{99}{\percent}$ for photons with energies $E > \gev{100}$~\cite{LHCfIJMPA}.

The multihit events that have more than one shower event in a single calorimeter
may appear due to pileup interactions in the same bunch crossing and then could
potentially cause a bias in the $\pt$ spectra.
However, considering the acceptance of the LHCf detector for inelastic
collisions $\sim 0.035$, the multihit probability due to the effects of pileup
is estimated only \SI{0.4}{\percent} and is therefore producing a negligible
effect. Detailed discussions of pileup effects and background events from
collisions between the beam and residual gas molecules in the beam tube can be
found in previous reports~\cite{LHCfIJMPA, LHCfphoton}.

Also beam divergence can cause a smeared beam spot at the TAN leading to a bias
in the measured $\pt$ spectra. The beam divergence at IP1 was
$\varepsilon/\beta^\ast = \SI{32}{\micro\radian}$~\cite{Jowett} for the three
fills mentioned, corresponding to a beam spot size at the TAN of roughly
$\sigma_\text{TAN} = \SI{4.5}{\milli\meter}$.
The effect of a non-zero beam spot size at the TAN is evaluated by comparing two
$\pt$ spectra predicted by toy MC simulations; one assuming a beam spot size of
zero and another assuming that the beam axis positions fluctuates following a
Gaussian distribution with $\sigma_\text{TAN} = \SI{4.5}{\milli\meter}$. The
$\pizero$ yield at $\pt = \gev{0.6}$ is found to increase by a factor 1.8 at
most. This effect is taken into account in the final results to the $\pt$
spectra.

\section{Monte Carlo simulations methodology}
\label{sec:mc_simulation}

MC simulations were performed in two steps: \\
(I) $p$--Pb interaction event generation at IP1 explained in
Sec.~\ref{sec:mc_signal} and (II) particle transport from IP1 to the LHCf
detector and consequent simulation of the response of the LHCf detector
(Sec.~\ref{sec:mc_detector}).

MC simulations which were then used for the validation of reconstruction
algorithms and cut criteria, and the estimation of systematic uncertainties
follow steps (I) and (II). These MC simulations are denoted as reference MC
simulations. On the other hand, MC simulations used only for comparisons with
measurement results in Sec.~\ref{sec:result} are limited to step (I) only, since
the final $\pt$ spectra in Sec.~\ref{sec:result} are already corrected for
detector responses and eventual reconstruction bias.

\subsection{Signal modeling}\label{sec:mc_signal}

Whenever the impact parameter of proton and Pb is smaller than the radius of
each particle, soft-QCD induced events are produced. These $p$-Pb interactions
at $\snn = \tev{5.02}$ and the resulting flux of secondary particles emitted
into forward rapidity region with their kinematics are simulated using various
hadronic interaction models (\textsc{dpmjet 3.04}~\cite{DPM3}, \textsc{qgsjet}
II-03~\cite{QGS2}, and \textsc{epos 1.99}~\cite{EPOS}). \textsc{Dpmjet 3.04}
also takes into account the Fermi motion of the nucleons in Pb nucleus and the
Cronin effect~\cite{Cronin}. Fermi motion enhances the $\pizero$ yields at most
by \SI{5}{\percent} in the LHCf $\pt$ covered range $\pt < \gev{0.6}$,
while the Cronin effect is not significant in this $\pt$ range.

On the other hand, when the impact parameter is larger than the overlapping
radii of each particle, so-called ultra-peripheral collisions (UPCs) can occur.
In UPCs virtual photons are emitted by the relativistic Pb nucleus which can
then collide with the proton beam~\cite{UPC}. The energy spectrum of these
virtual photons follows the Weizs\"{a}cker-Williams approximation~\cite{WW}.
The \textsc{sophia}~\cite{SOPHIA} MC generator is used to simulate the
photon--proton interaction in the rest frame of the proton and then the
secondary particles generated by \textsc{sophia} are boosted along the proton
beam.
For heavy nuclei with the radius $R_\text{A}$, the virtuality of the photon
$|q^2|<(\hslash c / R_\text{A})^2$ can be neglected and the photons are regarded
as real photon in the simulation in this analysis.

In these MC simulations, $\pizero$s from short-lived particles that decay within
\SI{1}{\meter} from IP1, mostly $\eta$ mesons decaying into $3\pizero$
($\lesssim \SI{10}{\percent}$ relative to all $\pizero$s), are also accounted
for consistently with the treatment of the experimental data.

\subsection{Simulation of particle transport from IP1 to the LHCf detector and
of the detector response}\label{sec:mc_detector}

The generated secondary particles are transported in the beam pipe from IP1 to
the TAN, taking into account the bending of charged particles' trajectory by the
Q1 quadrupole and the D1 beam separation dipole, particle decays, and particle
interactions with the beam pipe wall and the Y-shaped beam-vacuum-transition
chamber made of copper.

Finally the showers produced in the LHCf detector by the particles arriving at
the TAN and the detector response are simulated with the {\sc cosmos} and {\sc
epics} libraries~\cite{EPICS}. The detector position survey data and random
fluctuations due to electrical noise are taken into account. The simulations of
the LHCf detector are tuned to the test beam data taken at SPS, CERN in 2007 and
2012~\cite{SPS2007, SPSneutron}.

%
%
\section{Analysis framework}\label{sec:framework}

\subsection{$\pizero$ event reconstruction and
selection}\label{sec:reconstruction}

Since $\pizero$s decay into two photons very close to their point of creation at
IP1, each photon's direction is geometrically calculated using the impact
coordinates at the LHCf detector and the distance between IP1 and the detector
itself. Photon four-momentum is then derived by combining the photon's energy as
measured by the calorimeter with the previously obtained angle of emission.
Candidate $\pizero$s are selected from events where the invariant mass of the
two photons detected is within a narrow window around the $\pizero$ rest mass.

The $\pizero$ events are then classified in two categories: single-hit $\pizero$
and multihit $\pizero$ events.
The former is defined as having a single photon in each of the two calorimeter
towers only, while a multihit $\pizero$ event is defined as a single $\pizero$
accompanied by at least one additional background particle (usually a photon or
a neutron) in one of the two calorimeters.
In the analysis presented here, events having two particles within the same
calorimeter tower (multihit events) are rejected when the energy deposit of the
background particle is above a certain threshold~\cite{LHCfpi0}.
Mostly then, only single-hit $\pizero$ events are considered in this analysis.
The final inclusive production rates reported at the end are corrected for this
cut as described in Sec.~\ref{sec:pt_correction}.

Given the geometrical acceptance of the LHCf detector and to ensure a good event
reconstruction efficiency, the rapidity and $\pt$ range of the $\pizero$s are
limited to $-11.0 < \ylab < -8.9$ and $\pt < \gev{0.6}$, respectively. The
reconstructed invariant mass of the reference MC simulations peaks at $134.8 \pm
\mev{0.2}$ and reproduces well the measured data which peaks at $134.7 \pm
\mev{0.1}$, reproducing the $\pi^{0}$ mass.
The uncertainties given for the mass peaks are statistical only.

Standard reconstruction algorithms used for this analysis are described in
Ref.~\cite{LHCfpi0, LHCfIJMPA} and the $\pizero$ event selection criteria that
are applied prior to the reconstruction of the $\pizero$ kinematics are
summarized in Tab.~\ref{tbl:eventselection}. Systematic uncertainties are
discussed in Sec.~\ref{sec:systerror}.

\begin{table}
  \begin{center}
	\caption{Summary of the criteria for the selection of the $\pizero$ sample.}
	\begin{tabular}{c c}
      \hline
      \hline
      Incident position & within \SI{2}{\milli\meter} from the edge of
      calorimeter \\
      Energy threshold  & $E_\text{photon} > \gev{100}$   \\
      Number of hits    & Single-hit in each calorimeter  \\
      PID               & Photon like in each calorimeter \\
      \hline
      \hline
    \end{tabular}
    \label{tbl:eventselection}
  \end{center}
\end{table}

\subsection{Background subtraction}\label{sec:bg_subtraction}

Background contamination of the $\pizero$ events from hadronic events and the
causal coincidence of two photons not originated from the decay of a single
$\pizero$ are taken into account by subtracting the relevant contribution using
a sideband method~\cite{LHCfpi0}.

\begin{figure}[htbp]
  \centering
  \includegraphics[width=6cm, keepaspectratio]{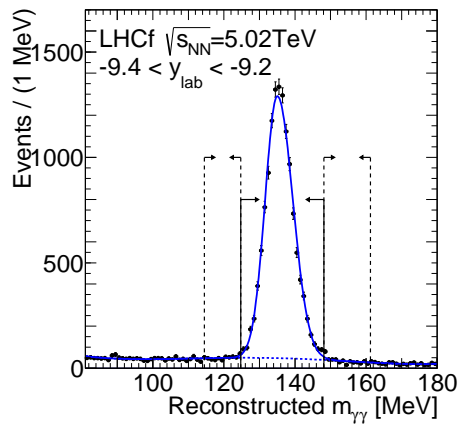}
  \caption{(color online). Reconstructed invariant mass distribution within the
  rapidity range from $-9.4$ to $-9.2$. The solid curve shows the best-fit composite
  physics model to the invariant mass distribution. The dashed curve indicates the
  background component. Solid and dashed lines indicate the signal and
  background windows, respectively.}
  \label{fig:mass-fit}
\end{figure}

Figure~\ref{fig:mass-fit} shows the reconstructed two-photon invariant mass
distribution of the experimental data in the rapidity range $-9.4 < \ylab <
-9.2$. The sharp peak around \mev{135} is owing to $\pizero$ events.
The solid curve indicates the best fit of a composite physics model to the
invariant mass distribution of the data; an asymmetric Gaussian distribution for
the signal component and a third order Chebyshev polynomial function for the
background component (dashed curve).
The signal window is defined as the invariant mass region within the two solid
arrows shown in Fig.~\ref{fig:mass-fit}, where the lower and upper limits are
given by $\hat{m}-3\sigma_l$ and $\hat{m}+3\sigma_u$, respectively.
$\hat{m}$ denotes the expected mean, and $\sigma_l$ and $\sigma_u$ denote
1\,$\sigma$ deviations for lower and upper side of the signal component,
respectively. The signal-rich rapidity--$\pt$ distributions are obtained by the
events contained inside of the signal window. The remaining contribution of
background events in the signal window is eliminated using the rapidity--$\pt$
distributions obtained from the background window, constructed from the two
sideband regions, [$\hat{m}-6\sigma_l$, $\hat{m}-3\sigma_l$] and
[$\hat{m}+3\sigma_u$, $\hat{m}+6\sigma_u$], that are defined as the invariant
mass regions within the dashed arrows in Fig.~\ref{fig:mass-fit}.

\subsection{Corrections to the $\pt$ spectra}\label{sec:pt_correction}

The raw rapidity--$\pt$ distributions must be corrected for (1) the
reconstruction inefficiency and the smearing caused by finite position and
energy resolutions, (2) geometrical acceptance and branching ratio of $\pizero$
decay, and (3) the loss due to multihit $\pizero$.

First, an iterative Bayesian method~\cite{Dagostini} is used to simultaneously
correct for both the reconstruction inefficiency and the smearing. The unfolding
procedure for the data is found in paper~\cite{LHCfpi0}.

Next, a correction for the limited aperture of the LHCf calorimeters must be
applied. The correction is derived from the rapidity--$\pt$ phase space.
The determination of the correction coefficients follow the same method used in
the $\pizero$ event analysis in $p$--$p$ collisions at $\sqs =
\tev{7}$~\cite{LHCfpi0}.

Finally, the loss of multihit $\pizero$ events, briefly mentioned in
Sec.~\ref{sec:reconstruction}, is corrected for using the MC event generators.
A range of ratios of multihit plus single-hit to single-hit $\pizero$ events is
estimated using three different hadronic interaction models ({\sc dpmjet 3.04},
{\sc qgsjet} II-03, and {\sc epos 1.99}) for each rapidity and $\pt$ range. The
observed $\pt$ spectra are then multiplied by the average of these ratios and
the contribution to the systematic uncertainty is derived from the observed
variations among the interaction models. The estimated range of the flux of
multihit $\pizero$ events lies within a range
\SI{0}{\percent}--\SI{10}{\percent} of the flux of single-hit $\pizero$ events.
The single-hit $\pizero$ spectra are then corrected to represent the inclusive
$\pizero$ production spectra.

All the procedures were verified using the reference MC simulations mentioned in
Sec.~\ref{sec:mc_simulation}.

%
%
\section{Systematic uncertainties}\label{sec:systerror}

We follow the same approach to estimate the systematic uncertainties as in
Ref.~\cite{LHCfpi0}.
Since systematic uncertainties on particle identification, single-hit selection,
and position-dependent corrections for both shower leakage and light yield of
the calorimeter are independent of the beam energy and type, the systematic
errors are taken directly from Ref.~\cite{LHCfpi0}.

Other terms deriving from the energy scale, beam axis offset, and luminosity are
updated consistently to the current understanding of the LHCf detector and the
beam configuration in $p$--Pb collisions. These terms are discussed in the
following subsections.
Table~\ref{tbl:systematicerror} summarises the systematic uncertainties for the
analysis.

\subsection{Energy scale}\label{sec:syst_energy}

The uncertainty on the measured photon energy was investigated in the beam test
at SPS~\cite{SPS2007} and also by performing a calibration with a radioactive
source~\cite{LHCfJINST}.
The estimated uncertainty on the photon energy from these tests is valued at
\SI{3.5}{\percent}.
This uncertainty is dominated by the conversion factors that relate measured
charge to deposited energy~\cite{SPS2007} and in fact, the reconstructed
invariant mass of two photons reproduces the $\pizero$ rest mass within the
uncertainty of \SI{3.5}{\percent} as shown in Fig.~\ref{fig:mass-fit}.

The systematic shift of bin contents due to energy scale uncertainties is
estimated using two different $\pt$ spectra in which the photon energy is
artificially scaled to the two extremes ($\pm \SI{3.5}{\percent}$). The ratios
of the two extreme spectra to the non-scaled spectrum are assigned as systematic
shifts of bin contents for each bin.

\subsection{Beam axis offset}\label{sec:syst_beamcenter}

The projected position of the proton beam axis on the LHCf detector (beam
centre) varies from Fill to Fill owing to the beam configuration, beam
transverse position and crossing angles, at IP1. The beam centre on the LHCf
detector can be determined by two methods; first we use the distribution of
incident particle positions as measured by the LHCf detector and second we also
use the information from the beam position monitors (BPMSW) installed $\pm
\SI{21}{\meter}$ from IP1~\cite{BPM}.

From analysis results in $p$--$p$ collisions at $\sqs=\tev{7}$, the beam centre
positions obtained by the two methods applied to LHC Fills 1089--1134 were found
to be consistent within \SI{1}{\milli\meter}.
The systematic shifts to the $\pt$ spectra are then evaluated by taking the
ratio of spectra with the beam centre displaced by $\pm\SI{1}{\milli\meter}$ to
spectra with no displacement present.
The fluctuations of the beam centre position modify the  $\pt$ spectra by
\SI{5}{\percent}--\SI{20}{\percent} depending on the rapidity range.

\subsection{Luminosity}\label{sec:syst_luminosity}

The luminosity value used for the analysis is derived from on the online
information provided by the ATLAS experiment. Since there is currently no robust
estimation on the luminosity error by the ATLAS experiment, we assign a
conservative $\pm \SI{20}{\percent}$ to the uncertainty. A more precise
estimation of the luminosity will be reported in future by the ATLAS
collaboration.

\begin{table}
  \centering
    \caption{Summary of the systematic uncertainties. Numerical values indicate the
    maximum variation of bin contents in the $\pt$ spectra.}
    \begin{tabular}{c c}
      \hline
      \hline
	  Energy scale            & \SI{5}{\percent}--\SI{20}{\percent} \\
      Particle identification & \SI{0}{\percent}--\SI{20}{\percent} \\
      Offset of beam axis     & \SI{5}{\percent}--\SI{20}{\percent} \\
      Single-hit selection    & \SI{3}{\percent} \\
      Position-dependent correction & \SI{5}{\percent}--\SI{30}{\percent} \\
      Luminosity              & \SI{20}{\percent} \\
      \hline
      \hline
    \end{tabular}
    \label{tbl:systematicerror}
\end{table}

\section{Results and discussion}\label{sec:result}

\subsection{The QCD induced transverse momentum distribution}
\label{sec:pt_withoutupc}

\begin{figure*}[htbp]
  \begin{center}
  \includegraphics[width=16cm, keepaspectratio]{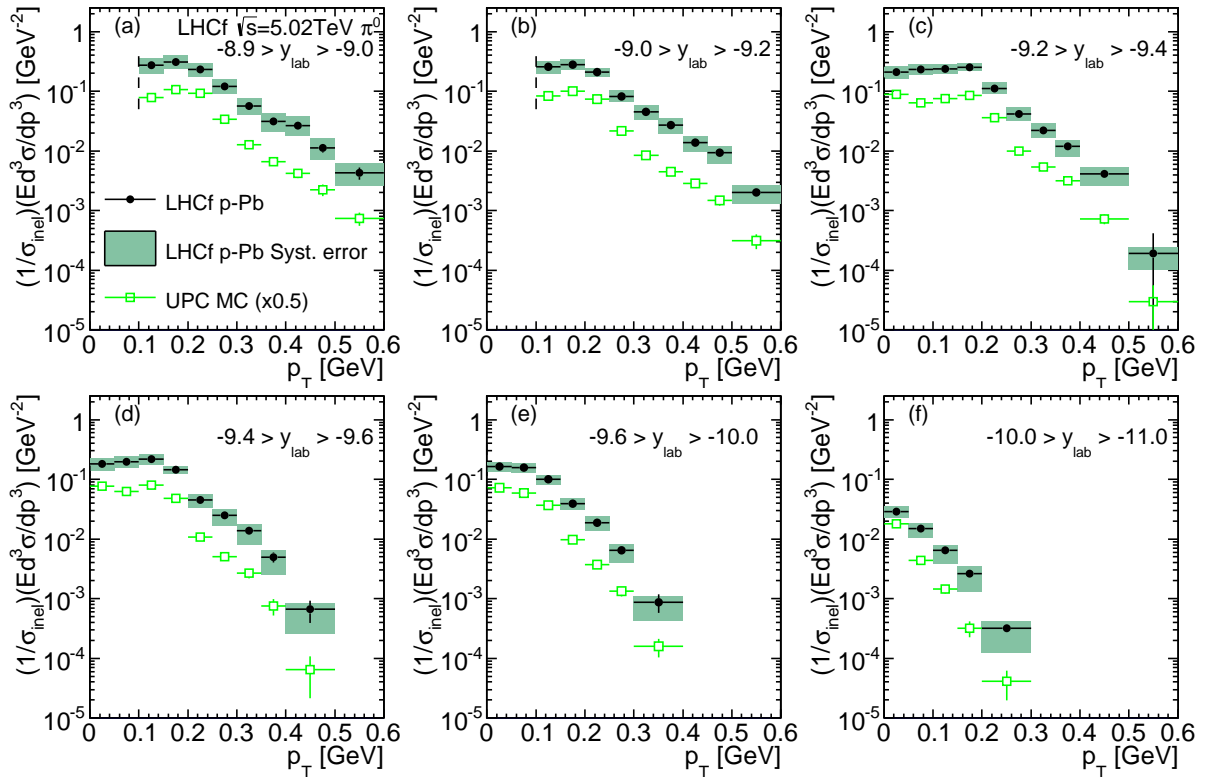}
  \caption{(color online). Experimental $\pt$ spectra of the LHCf detector
  (filled circles). Error bars and shaded rectangles indicate the statistical
  and systematic uncertainties respectively. The open squares indicate the
  estimated contribution from UPCs.}
  \label{fig:pt_withupc}
  \end{center}
\end{figure*}

The $\pt$ spectra obtained from the data analysed are presented in
Fig.~\ref{fig:pt_withupc}. The spectra are categorized into six ranges of
rapidity $\ylab$: [-9.0, -8.9], [-9.2, -9.0], [-9.4, -9.2], [-9.6, -9.4],
[-10.0, -9.6], and [-11.0, -10.0]. The spectra have all the corrections
discussed in Sec.~\ref{sec:pt_correction} applied. The inclusive $\pizero$
production rate is given as
\begin{equation}
    \frac{1}{\sigma^\text{pPb}_\text{inel}} E\frac{d^{3}\sigma^\text{pPb}}{dp^3}
    =
    \frac{1}{N^\text{pPb}_\text{inel}}
    \frac{d^2 N^\text{pPb}(\pt, y)}{2\pi \pt d\pt dy}.
\end{equation}
\noindent where $\sigma^\text{pPb}_\text{inel}$ is the inelastic cross section
for $p$--Pb collisions at $\snn = \tev{5.02}$ and $E d^{3}\sigma^\text{pPb} /
dp^3$ is the inclusive cross section of $\pizero$ production.
The number of inelastic $p$--Pb collisions, $N^\text{pPb}_\text{inel}$, used for
normalizing the production rates of Fig.~\ref{fig:pt_withupc} is calculated from
$N^\text{pPb}_\text{inel}$ = $\sigma^\text{pPb}_\text{inel} \int {\cal L} dt$,
assuming an inelastic $p$--Pb cross section $\sigma^\text{pPb}_\text{inel} =
\SI{2.11}{\barn}$. The value for $\sigma^\text{pPb}_\text{inel}$ is derived
from the inelastic $p$--$p$ cross section $\sigma^\text{pp}_\text{inel}$ and the
Glauber multiple collision model~\cite{Glauber, dEnterria}.
Using the integrated luminosities shown in Sec.~\ref{sec:data},
$N^\text{pPb}_\text{inel}$ is 9.33$\times$10$^{7}$. $d^2N^\text{pPb} (\pt, y)$
is the number of $\pizero$s detected in the transverse momentum interval
($d\pt$) and the rapidity interval ($dy$) with all corrections applied.

In Fig.~\ref{fig:pt_withupc}, the filled circles represent the data from the
LHCf experiment. The error bars and shaded rectangles indicate the 1 standard
deviation statistical and total systematic uncertainties respectively.
The total systematic uncertainties are given by adding all uncertainty terms
except the one for luminosity in quadrature. The vertical dashed lines shown for
the rapidity ranges greater than $-9.2$ indicate the $\pt$ threshold of the LHCf
detector due to the photon energy threshold and the geometrical acceptance of
the detector.
The contribution from UPCs is presented as open squares (normalized to $1/2$ for
visibility). This UPC contribution is estimated with the MC simulations
introduced in Sec.~\ref{sec:mc_signal} using the UPC cross section
from~\cite{LHCfUPC}.

To obtain the soft-QCD component, the UPC contribution must be subtracted from
the measured $\pt$ spectra. This is achieved by simply subtracting, point by
point, the UPC induced $\pt$ spectra (open squares in Fig.~\ref{fig:pt_withupc})
from the total $\pt$ spectra (filled circles in Fig.~\ref{fig:pt_withupc}).
Uncertainties in the subtracted results are obtained by adding the statistical
and systematic errors in quadrature.
The theoretical uncertainty on the UPC estimation $\pm \SI{5}{\percent}$ derives
mainly from the knowledge of the virtual photon flux given by the relativistic
Pb nucleus and of the virtual photon-proton cross section~\cite{LHCfUPC}.
The inclusive production rates of $\pizero$s measured by LHCf after the
subtraction of the UPC component are summarized in Appendix A.

Figure~\ref{fig:pt_withoutupc_simple} shows the LHCf data $\pt$ spectra after
subtraction of the UPC component (filled circles). The size of the error bars
correspond to \SI{68}{\percent} confidence intervals (including both statistical
and systematic uncertainties).
The $\pt$ spectra in $p$--Pb collisions at \tev{5.02} predicted by the hadronic
interaction models, {\sc dpmjet 3.04} (solid line, red), {\sc qgsjet} II-03
(dashed line, blue), and {\sc epos 1.99} (dotted line, magenta), are also shown
in the same figure for comparison. Predictions by the three hadronic interaction
models do not include the UPC component. The experimental $\pt$ spectra are
corrected for detector response, event selection and geometrical acceptance
efficiencies, so that the $\pt$ spectra of the interaction models can be
compared directly to the experimental spectra.

In Fig.~\ref{fig:pt_withoutupc_simple}, among the predictions given by the
hadronic interaction models tested here, {\sc dpmjet 3.04} and {\sc epos 1.99}
show a good overall agreement with the LHCf measurements. However {\sc qgsjet}
II-03 predicts softer $\pt$ spectra than the LHCf measurements and the other two
hadronic interaction models. Similar features of these hadronic interaction
models are also seen in $p$--$p$ collisions at $\sqs = \tev{7}$~\cite{LHCfpi0}.

In Fig.~\ref{fig:pt_withoutupc_simple}, the $\pt$ spectra in $p$--$p$ collisions
at $\sqs = \tev{5.02}$ are also added and will be useful for the derivation of
the nuclear modification factor described later in Sec.~\ref{sec:nmf_result}.
These spectra are multiplied by a factor 5 for visibility. The derivation of the
$\pt$ spectra in $p$--$p$ collisions at $\sqs = \tev{5.02}$ is explained in
detail in Appendix B.

\begin{figure*}[htbp]
  \begin{center}
  \includegraphics[width=16cm, keepaspectratio]{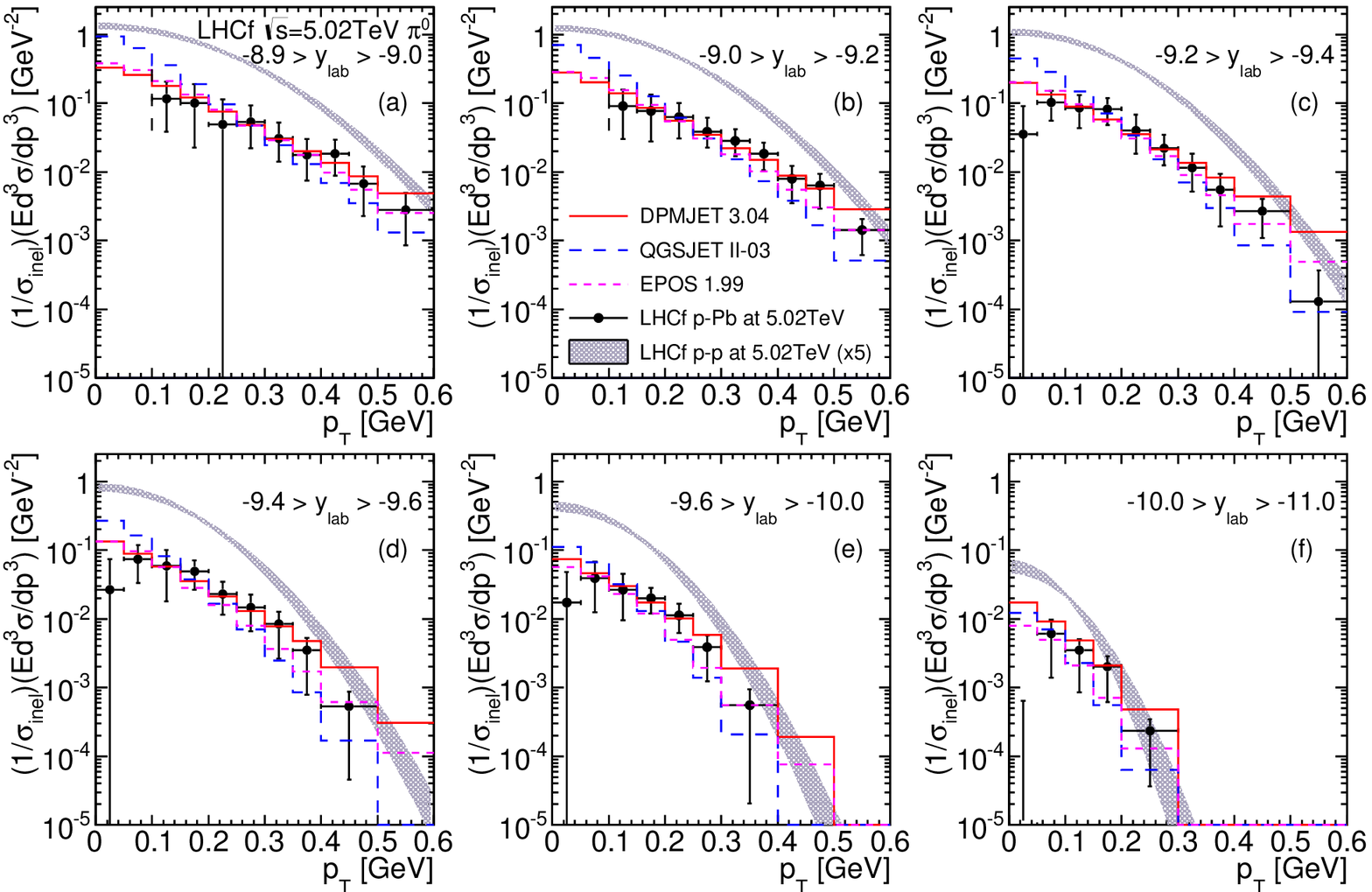}
  \caption{(color online). Experimental $\pt$ spectra measured by LHCf after the
  subtraction of the UPC component (filled circles). Error bars indicate the
  total uncertainties incorporating both statistical and systematic
  uncertainties. Hadronic interaction models predictions and derived 
  spectra for $p$--$p$ collisions at \tev{5.02} are also shown (see text for
  details).}
  \label{fig:pt_withoutupc_simple}
  \end{center}
\end{figure*}

\subsection{Average transverse momentum.}\label{sec:average_pt}

The average transverse momentum, $\langle\pt\rangle$, can be obtained by fitting
an empirical function to the $\pt$ spectra in
Fig.~\ref{fig:pt_withoutupc_simple} for each rapidity range.
Two distributions to parametrize the $\pt$ spectra were chosen among the several
proposed in literature:
an exponential and a Gaussian.
Detailed descriptions of the parametrization and derivation of
$\langle\pt\rangle$ can be found in Ref.~\cite{LHCfpi0}.

\begin{figure}[htbp]
  \begin{center}
  \includegraphics[width=8.0cm, keepaspectratio]{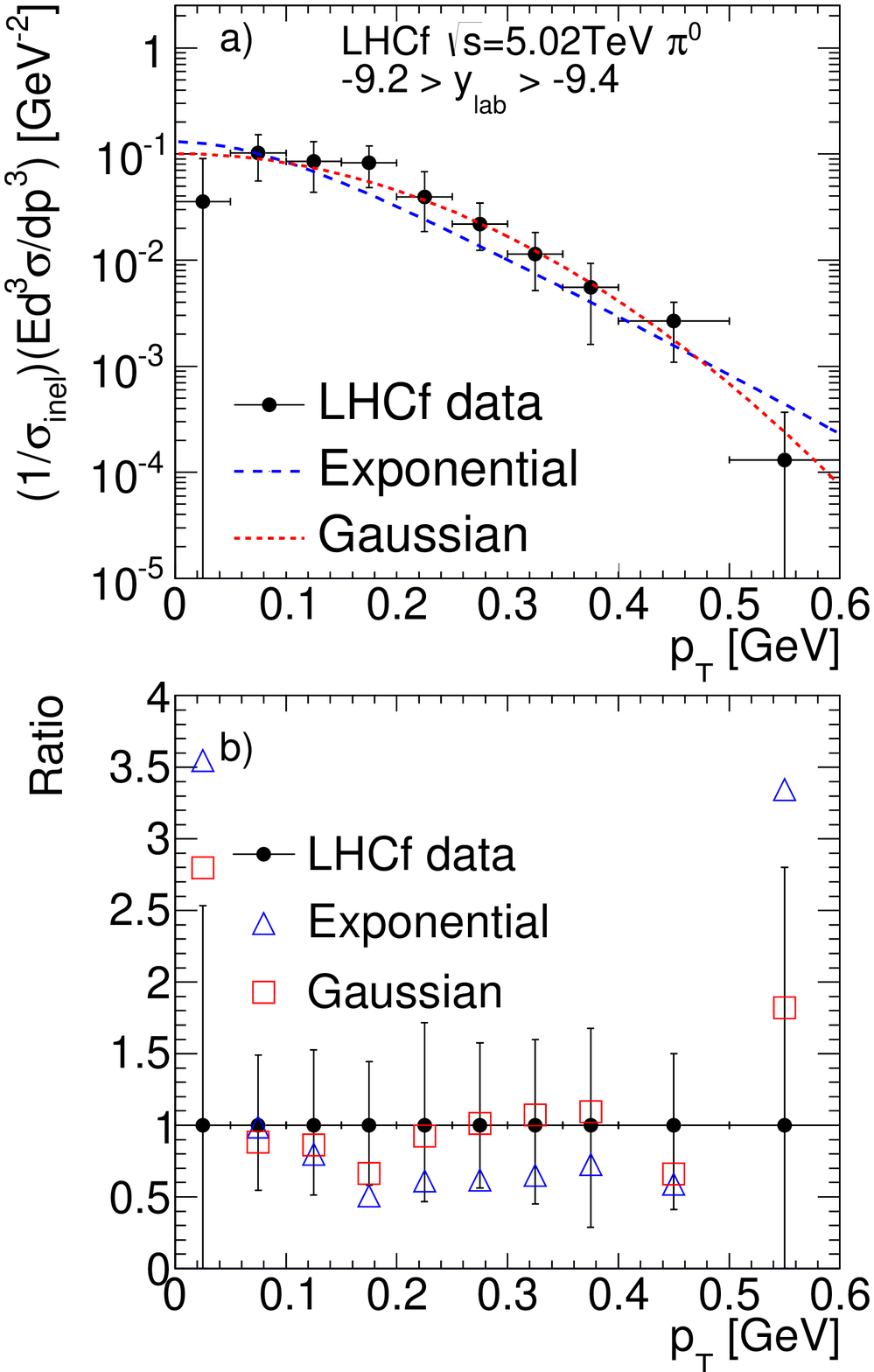}
  \caption{(color online). (a) Upper: experimental $\pt$ spectra (filled circles
  and error bars), the best-fit exponential (dashed curve) and Gaussian
  distributions (dotted curve). (b) Bottom: ratios of the best-fit exponential
  or Gaussian distribution to the experimental data (open triangles or open
  boxes) with statistical and systematic uncertainties (filled circles and error
  bars.)}
  \label{fig:pt_fit}
  \end{center}
\end{figure}

For example, the upper panel in Fig.~\ref{fig:pt_fit} shows the experimental
$\pt$ spectra (filled circles) and the best fit with the exponential
(dashed curve, blue) and with the Gaussian distribution
(dotted curve, red) in the rapidity range $-9.2 > \ylab > -9.4$. The bottom
panel in Fig.~\ref{fig:pt_fit} shows best-fit ratio to the experimental data;
exponential (blue open triangles) and Gaussian distributions (red open circles).
Error bars indicate the statistical and systematic uncertainties in the both
panels.

On the other hand, $\langle\pt\rangle$ can be simply estimated by numerically
integrating the $\pt$ spectra in Fig.~\ref{fig:pt_withoutupc_simple}.
In this approach, $\langle\pt\rangle$ is obtained over the rapidity range $-9.2
> \ylab > -10.0$ where the $\pt$ spectra are available down to \gev{0}. The data
in the rapidity range $-10.0 > \ylab > -11.0$ is not used here, since the bin
content in $\gev{0}< \pt < \gev{0.05}$ is negative due to the subtraction of
UPCs.
Although the interval for the numerical integration is bounded from above by
$\pt^\text{upper}$, the high $\pt$ tail contribution to $\langle\pt\rangle$ is
negligibly small.

The values of $\langle\pt\rangle$ obtained by the three methods are in good
agreement within the uncertainties. The specific values of $\langle\pt\rangle$
for this paper, $\langle\pt\rangle_\text{LHCf}$, are defined as follows.
\noindent For the rapidity range $-9.2>\ylab>-10.0$, the values of
$\langle\pt\rangle_\text{LHCf}$ are taken from the weighted average of
$\langle\pt\rangle$ from the exponential fit, the Gaussian fit, and the
numerical integration. The uncertainty related to a possible bias of the
$\langle\pt\rangle$ extraction methods is derived from the largest value among
the three methods. For the other rapidity ranges to where the numerical
integration is not applicable, the weighted mean and the uncertainty are
obtained following the same method above but using only the exponential and the
Gaussian fit.
Best-fit results for the three approaches mentioned above and the values of
$\langle\pt\rangle_\text{LHCf}$ are summarised in Table~\ref{table:average_pt}.

\begin{figure}[htbp]
  \begin{center}
  \includegraphics[width=7cm, keepaspectratio]{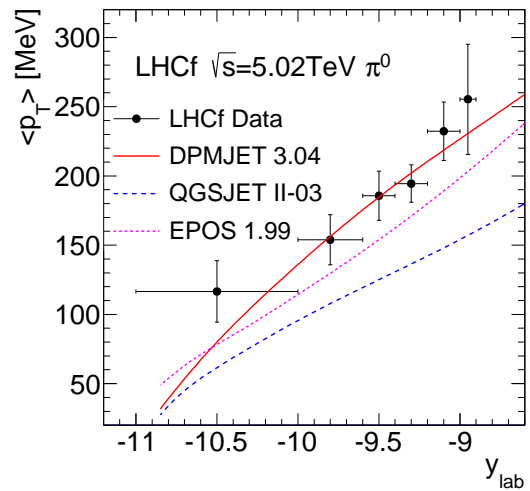}
  \caption{(color online). Average $\pt$ as a function of rapidity $\ylab$.
  Filled circles indicate the LHCf data. The predictions of hadronic interaction
  models are shown (solid curve {\sc dpmjet} 3.04, dashed curve {\sc qgsjet}
  II-03, and dotted curve {\sc epos} 1.99).}
  \label{fig:average_pt}
  \end{center}
\end{figure}

Figure~\ref{fig:average_pt} shows the $\langle\pt\rangle_\text{LHCf}$ and the
predictions by hadronic interaction models as a function of the rapidity
$\ylab$. The average $\pt$ of the hadronic interaction models is calculated by
numerical integration. {\sc dpmjet} 3.04 reproduces quite well the LHCf
measurements $\langle\pt\rangle_\text{LHCf}$, while {\sc epos} 1.99 is slightly
softer than both {\sc dpmjet} 3.04 and the LHCf measurements. {\sc qgsjet} II-03
shows the smallest $\langle\pt\rangle$ among the three models and the LHCf
measurements. These tendencies are also found in
Fig.~\ref{fig:pt_withoutupc_simple}.

\begin{widetext}
\begin{table*}[htbp]
    \begin{center}
    \caption{The average $\pizero$ transverse momenta for the rapidity range
    $-8.9>\ylab>-11.0$ estimated by the three approaches (exponential, Gaussian,
    and numerical integration). Combined results using the three approaches are
    denoted as LHCf results. \label{table:average_pt}}
    \begin{tabular}{l    r c c  r c c  c c c  c c} \hline\hline
    \multicolumn{1}{l }{~} &
    \multicolumn{3}{c}{Exponential fit}       &
    \multicolumn{3}{c}{Gaussian fit}          &
    \multicolumn{3}{c}{Numerical integration} &
    \multicolumn{2}{c}{LHCf analysis}         \\
    Rapidity
    & $\chi^2$ (dof)     & $\langle\pt\rangle$ & Stat. error
    & $\chi^2$ (dof)     & $\langle\pt\rangle$ & Stat. error
    & $\pt^\text{upper}$ & $\langle\pt\rangle$ & Stat. error
    & $\langle\pt\rangle_\text{LHCf}$          & Syst. error  \\
    &                    & [MeV]               & [MeV]
    &                    & [MeV]               & [MeV]
    & [GeV]              & [MeV]               & [MeV]
    & [MeV]              & [MeV]                              \\
    \hline
    $[$-8.9, -9.0$]$  &   0.9 (7)  & 249.1 & 36.8
                      &   0.8 (7)  & 258.5 & 27.9
                      &            &       &
                                   & 255.3 & 36.8         \\
    $[$-9.0, -9.2$]$  &   2.0 (7)  & 221.1 & 20.4
                      &   0.5 (7)  & 239.5 & 16.6
                      &            &       &
                                   & 232.3 & 20.4         \\
    $[$-9.2, -9.4$]$  &   7.6 (8)  & 188.7 & 13.4
                      &   2.7 (8)  & 196.4 &  8.6
                      &   0.6      & 193.3 & 13.2         
                                   & 194.4 & 13.4         \\
    $[$-9.4, -9.6$]$  &   4.8 (6)  & 181.8 & 16.3
                      &   1.6 (6)  & 187.0 & 12.7
                      &   0.5      & 184.6 & 14.9         
                                   & 185.7 & 16.3         \\
    $[$-9.6, -10.0$]$ &   3.7 (5)  & 153.0 & 16.3
                      &   1.5 (5)  & 153.7 & 12.3
                      &   0.4      & 152.2 & 13.9
                                   & 153.9 & 16.3         \\
    $[$-10.0,-11.0$]$ &  <0.1 (2)  & 115.1 & 22.2
                      &  <0.1 (2)  & 117.5 & 17.5
                      &            &       &
                                   & 116.6 & 22.2         \\
    \hline\hline

    \end{tabular}
    \end{center}
\end{table*}
\end{widetext}

\subsection{Nuclear modification factor}\label{sec:nmf_result}

Finally the nuclear modification factor $\nmf$ is derived. This factor
quantifies the $\pt$ spectra modification caused by nuclear effects in $p$--Pb
collisions. The nuclear modification factor is defined as
\begin{equation}
	\nmf \equiv
	\frac{\sigma^\textrm{pp}_\textrm{inel}}
	{\langle N_\textrm{coll} \rangle \sigma^\textrm{pPb}_\textrm{inel}}
	\frac{Ed^3\sigma^\textrm{pPb}/dp^3}{Ed^3\sigma^\textrm{pp}/dp^3},
    \label{eq:nmf}
\end{equation}
\noindent where $E d^{3}\sigma^\text{pPb} / dp^3$ and $E d^{3}\sigma^\text{pp} /
dp^3$ are the inclusive cross sections of $\pizero$ production in $p$--Pb and
$p$--$p$ collisions at \tev{5.02} respectively.
The UPC component is already subtracted in $\sigma^\text{pPb}$.
The uncertainty on $\sigma^\text{pPb}_\text{inel}$ is estimated to be $\pm
\SI{5}{\percent}$ by comparing the $\sigma^\text{pPb}_\text{inel}$ value with
other calculations and experimental results presented in~\cite{CMSxsec,
ALICExsec}. The average number of binary nucleon--nucleon collisions in a
$p$--Pb collision, $\langle N_\textrm{coll} \rangle = 6.9$, is obtained from MC
simulations using the Glauber model~\cite{dEnterria}. The uncertainty in
$\sigma^\textrm{pp}_\textrm{inel}/\langle N_\textrm{coll} \rangle$ is estimated
by varying the parameters in the calculation with the Glauber model~\cite{ALICE,
ALICErapidity} (where the cancellation of the uncertainties in
$\sigma^\textrm{pp}_\textrm{inel}$ and $\langle N_\textrm{coll} \rangle$ is
taken into account) and is of the order of $\pm\SI{3.5}{\percent}$ . Finally the
quadratic sum of the uncertainties in $\sigma^\text{pPb}_\text{inel}$ and
$\sigma^\textrm{pp}_\textrm{inel}/\langle N_\textrm{coll} \rangle$ is added to
$\nmf$.

Since there is no data at $\sqs=\tev{5.02}$ for $p$--$p$ collisions $E
d^{3}\sigma^\text{pp} / dp^3$ is derived by scaling the $\pt$ spectra taken in
$p$--$p$ collisions at \tev{7} and \tev{2.76}.
The derivation follows three steps. First (I) the $\langle\pt\rangle$ values at
\tev{5.02} are estimated by interpolating the $\langle\pt\rangle$ values at
\tev{7} and \tev{2.76}, assuming that the Feynman scaling of $\langle\pt\rangle$
is only a function of rapidity. Then (II) the absolute normalizations of the
$\pt$ spectra at \tev{5.02} are determined by applying the measured absolute
normalizations at
\tev{7} directly to those at \tev{5.02}. Finally (III) the $\pt$ spectra at \tev{5.02}
are derived assuming that the $\pt$ spectra follow a Gaussian distribution with
width $2\langle\pt\rangle/\sqrt{\pi}$ (obtained in step (I)) and using the
normalizations obtained in step (II). The rapidity shift $-0.465$ explained in
Sec.~\ref{sec:data} is also taken into account in the $\pt$ spectrum at
\tev{5.02}.
The details of the procedure are discussed in Appendix B.

Figure~\ref{fig:nmf_simple} shows the nuclear modification factors $\nmf$ from
the LHCf measurements and the predictions by hadronic interaction models {\sc
dpmjet 3.04} (red solid line), {\sc qgsjet} II-03 (blue dashed line), and {\sc
epos 1.99} (magenta dotted line).
The LHCf measurements, although with a large uncertainty which increases with
$\pt$ (mainly due to systematic uncertainties in $p$--Pb collisions at
$\tev{5.02}$), show a strong suppression with $\nmf$ equal 0.1 at $\pt \approx
\gev{0.1}$ rising to 0.3 at $\pt \approx \gev{0.6}$. All hadronic
interaction models predict small values of $\nmf \approx 0.1$, and they show an
overall good agreement with the LHCf measurements within the uncertainty.
Clearly other analyses which are more sensitive to exclusive $\pizero$ signals
are needed, for example diffractive dissociation, to investigate the reason for
this strong suppression. However the measured $\nmf$ dependency on $\pt$ and
rapidity may hint to an understanding of the break down of the $\pizero$
production mechanism.

\begin{figure*}[htbp]
  \begin{center}
  \includegraphics[width=16cm,keepaspectratio]{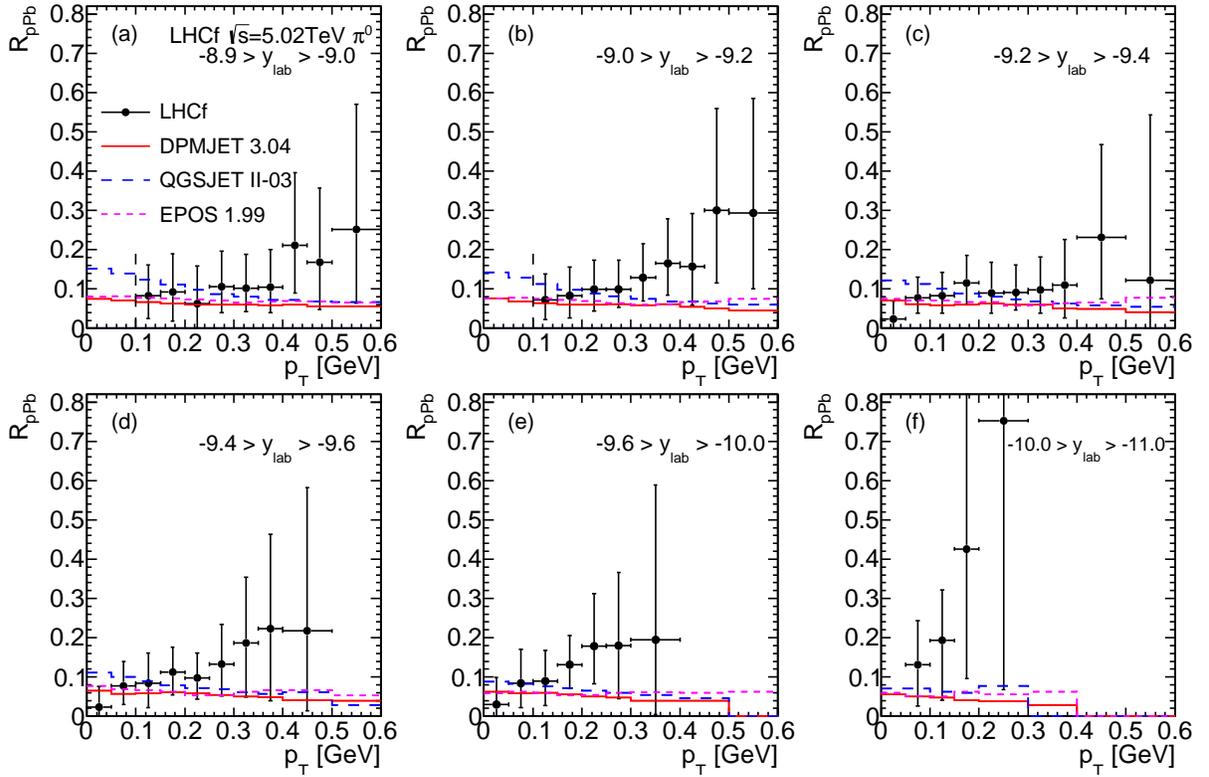}
  \caption{(color online). Nuclear modification factor for $\pizero$s.
  Filled circles indicate the factors obtained by the LHCf measurements. Error
  bars indicate the total uncertainties incorporating both statistical and
  systematic uncertainties. Other lines are the predictions by hadronic
  interaction models (see text in detail.)}
  \label{fig:nmf_simple}
  \end{center}
\end{figure*}

\section{Conclusions}\label{sec:conclusions}

The inclusive production of neutral pions in the rapidity range $-8.9 > \ylab >
-11.0$ has been measured by the LHCf experiment in $p$--Pb collisions at $\snn =
\tev{5.02}$ at the LHC in 2013.
Transverse momentum spectra of neutral pions measured by the LHCf detectors have
been compared with the predictions of several hadronic interaction models.
Among the hadronic interaction models tested in this paper, {\sc dpmjet} 3.04
and {\sc epos} 1.99 show the best overall agreement with the LHCf data in the
rapidity range $-8.9 > \ylab > -11.0$, while {\sc qgsjet} II-03 shows softer
$\pt$ spectra relative to the the LHCf data and the other two hadronic
interaction models. These tendencies are also recognized in the comparison on
the average $\pt$ distribution as a function of rapidity $\ylab$.

The nuclear modification factor, $\nmf$, derived from the LHCf measurements
indicates a strong suppression of the $\pizero$ production in the nuclear target
relative to those in the nucleon target. All hadronic interaction models present
an overall good agreement with the LHCf measurements within the uncertainty.

As a future prospect, additional analyses which are sensitive to exclusive
$\pizero$ spectra, are needed to reach a better understanding of this strong
suppression and its break down.

\section*{Acknowledgments}\label{sec:acknowledgements}

We thank the CERN staff and the ATLAS collaboration for their essential
contributions to the successful operation of LHCf. This work is partly supported
by Grant-in-Aid for Scientific research by MEXT of Japan, Grant-in-Aid for JSPS
Postdoctoral Fellow for Research Abroad, and the Grant-in-Aid for Nagoya
University GCOE "QFPU" from MEXT. This work is also supported by Istituto
Nazionale di Fisica Nucleare (INFN) in Italy.
A part of this work was performed using the computer resource provided by the
Institute for the Cosmic-Ray Research (University of Tokyo), CERN, and CNAF
(INFN).

\section*{Appendix A}\label{sec:appendix_a}

The inclusive production rates of $\pizero$s measured by LHCf after the
subtraction of the UPC component are summarized in
Tables~\ref{table:spectra_0}--~\ref{table:spectra_5}.

\begin{table*}[htbp]
    \begin{center}
    \caption{Production rate for the $\pizero$ production in the rapidity
    range $-8.9 > \ylab > -9.0$. \label{table:spectra_0}}
    \begin{tabular*}{14cm}{@{\extracolsep{\fill}}c c c}
    \hline
    \hline
    $\pt$ range $[$GeV$]$
    & Production rate $[$GeV$^{-2}]$
    & Syst+Stat uncertainty $[$GeV$^{-2}]$ \\
    \hline
$[$0.10, 0.15$]$ & 1.17$\times10^{-1}$ & -7.83$\times10^{-2}$, +8.91$\times10^{-2}$ \\
$[$0.15, 0.20$]$ & 1.00$\times10^{-1}$ & -7.75$\times10^{-2}$, +8.79$\times10^{-2}$ \\
$[$0.20, 0.25$]$ & 4.90$\times10^{-2}$ & -5.57$\times10^{-2}$, +6.43$\times10^{-2}$ \\
$[$0.25, 0.30$]$ & 5.37$\times10^{-2}$ & -3.18$\times10^{-2}$, +3.85$\times10^{-2}$ \\
$[$0.30, 0.35$]$ & 3.09$\times10^{-2}$ & -1.69$\times10^{-2}$, +2.11$\times10^{-2}$ \\
$[$0.35, 0.40$]$ & 1.76$\times10^{-2}$ & -1.02$\times10^{-2}$, +1.24$\times10^{-2}$ \\
$[$0.40, 0.45$]$ & 1.82$\times10^{-2}$ & -9.42$\times10^{-3}$, +1.12$\times10^{-2}$ \\
$[$0.45, 0.50$]$ & 6.77$\times10^{-3}$ & -4.49$\times10^{-3}$, +5.24$\times10^{-3}$ \\
$[$0.50, 0.60$]$ & 2.78$\times10^{-3}$ & -1.93$\times10^{-3}$, +2.20$\times10^{-3}$ \\
    \hline
    \hline
    \end{tabular*}
    \end{center}
\end{table*}

\begin{table*}[htbp]
    \begin{center}
    \caption{Production rate for the $\pizero$ production in the rapidity
    range $-9.0 > \ylab > -9.2$. \label{table:spectra_1}}
    \begin{tabular*}{14cm}{@{\extracolsep{\fill}}c c c}
    \hline
    \hline
    $\pt$ range $[$GeV$]$
    & Production rate $[$GeV$^{-2}]$
    & Syst+Stat uncertainty $[$GeV$^{-2}]$ \\
    \hline
$[$0.10, 0.15$]$ & 9.06$\times10^{-2}$ & -6.05$\times10^{-2}$, +6.61$\times10^{-2}$ \\
$[$0.15, 0.20$]$ & 7.71$\times10^{-2}$ & -4.92$\times10^{-2}$, +5.54$\times10^{-2}$ \\
$[$0.20, 0.25$]$ & 6.28$\times10^{-2}$ & -3.23$\times10^{-2}$, +3.84$\times10^{-2}$ \\
$[$0.25, 0.30$]$ & 3.83$\times10^{-2}$ & -1.58$\times10^{-2}$, +2.31$\times10^{-2}$ \\
$[$0.30, 0.35$]$ & 2.81$\times10^{-2}$ & -1.13$\times10^{-2}$, +1.36$\times10^{-2}$ \\
$[$0.35, 0.40$]$ & 1.82$\times10^{-2}$ & -7.76$\times10^{-3}$, +8.27$\times10^{-3}$ \\
$[$0.40, 0.45$]$ & 7.98$\times10^{-3}$ & -4.49$\times10^{-3}$, +4.33$\times10^{-3}$ \\
$[$0.45, 0.50$]$ & 6.37$\times10^{-3}$ & -3.43$\times10^{-3}$, +3.05$\times10^{-3}$ \\
$[$0.50, 0.60$]$ & 1.41$\times10^{-3}$ & -8.01$\times10^{-4}$, +6.56$\times10^{-4}$ \\
    \hline
    \hline
    \end{tabular*}
    \end{center}
\end{table*}

\begin{table*}[htbp]
    \begin{center}
    \caption{Production rate for the $\pizero$ production in the rapidity
    range $-9.2 > \ylab > -9.4$. \label{table:spectra_2}}
    \begin{tabular*}{14cm}{@{\extracolsep{\fill}}c c c}
    \hline
    \hline
    $\pt$ range $[$GeV$]$
    & Production rate $[$GeV$^{-2}]$
    & Syst+Stat uncertainty $[$GeV$^{-2}]$ \\
    \hline
$[$0.00, 0.05$]$ & 3.56$\times10^{-2}$ & -5.14$\times10^{-2}$, +5.46$\times10^{-2}$ \\
$[$0.05, 0.10$]$ & 1.02$\times10^{-1}$ & -4.63$\times10^{-2}$, +4.99$\times10^{-2}$ \\
$[$0.10, 0.15$]$ & 8.53$\times10^{-2}$ & -4.15$\times10^{-2}$, +4.49$\times10^{-2}$ \\
$[$0.15, 0.20$]$ & 8.25$\times10^{-2}$ & -3.45$\times10^{-2}$, +3.67$\times10^{-2}$ \\
$[$0.20, 0.25$]$ & 3.96$\times10^{-2}$ & -2.11$\times10^{-2}$, +2.83$\times10^{-2}$ \\
$[$0.25, 0.30$]$ & 2.19$\times10^{-2}$ & -9.62$\times10^{-3}$, +1.26$\times10^{-2}$ \\
$[$0.30, 0.35$]$ & 1.14$\times10^{-2}$ & -6.25$\times10^{-3}$, +6.81$\times10^{-3}$ \\
$[$0.35, 0.40$]$ & 5.53$\times10^{-3}$ & -3.93$\times10^{-3}$, +3.75$\times10^{-3}$ \\
$[$0.40, 0.50$]$ & 2.67$\times10^{-3}$ & -1.57$\times10^{-3}$, +1.33$\times10^{-3}$ \\
$[$0.50, 0.60$]$ & 1.31$\times10^{-4}$ & -2.45$\times10^{-4}$, +2.36$\times10^{-4}$ \\
    \hline
    \hline
    \end{tabular*}
    \end{center}
\end{table*}

\begin{table*}[htbp]
    \begin{center}
	\caption{Production rate for the $\pizero$ production in the rapidity
    range $-9.4 > \ylab > -9.6$. \label{table:spectra_3}}
	\begin{tabular*}{14cm}{@{\extracolsep{\fill}}c c c}
    \hline
    \hline
    $\pt$ range $[$GeV$]$
    & Production rate $[$GeV$^{-2}]$
    & Syst+Stat uncertainty $[$GeV$^{-2}]$ \\
    \hline
$[$0.00, 0.05$]$ & 2.67$\times10^{-2}$ & -4.42$\times10^{-2}$, +4.64$\times10^{-2}$ \\
$[$0.05, 0.10$]$ & 7.42$\times10^{-2}$ & -4.11$\times10^{-2}$, +4.31$\times10^{-2}$ \\
$[$0.10, 0.15$]$ & 5.87$\times10^{-2}$ & -4.08$\times10^{-2}$, +4.19$\times10^{-2}$ \\
$[$0.15, 0.20$]$ & 4.93$\times10^{-2}$ & -2.26$\times10^{-2}$, +2.13$\times10^{-2}$ \\
$[$0.20, 0.25$]$ & 2.32$\times10^{-2}$ & -1.17$\times10^{-2}$, +1.16$\times10^{-2}$ \\
$[$0.25, 0.30$]$ & 1.47$\times10^{-2}$ & -8.10$\times10^{-3}$, +7.54$\times10^{-3}$ \\
$[$0.30, 0.35$]$ & 8.37$\times10^{-3}$ & -5.73$\times10^{-3}$, +4.31$\times10^{-3}$ \\
$[$0.35, 0.40$]$ & 3.47$\times10^{-3}$ & -2.68$\times10^{-3}$, +1.77$\times10^{-3}$ \\
$[$0.40, 0.50$]$ & 5.32$\times10^{-4}$ & -4.86$\times10^{-4}$, +3.39$\times10^{-4}$ \\
    \hline
    \hline
    \end{tabular*}
    \end{center}
\end{table*}

\begin{table*}[htbp]
    \begin{center}
    \caption{Production rate for the $\pizero$ production in the rapidity
    range $-9.6 > \ylab > -10.0$. \label{table:spectra_4}}
    \begin{tabular*}{14cm}{@{\extracolsep{\fill}}c c c}
    \hline
    \hline
    $\pt$ range $[$GeV$]$
    & Production rate $[$GeV$^{-2}]$
    & Syst+Stat uncertainty $[$GeV$^{-2}]$ \\
    \hline
$[$0.00, 0.05$]$ & 1.72$\times10^{-2}$ & -2.93$\times10^{-2}$, +3.07$\times10^{-2}$ \\
$[$0.05, 0.10$]$ & 3.93$\times10^{-2}$ & -2.69$\times10^{-2}$, +2.83$\times10^{-2}$ \\
$[$0.10, 0.15$]$ & 2.67$\times10^{-2}$ & -1.72$\times10^{-2}$, +1.81$\times10^{-2}$ \\
$[$0.15, 0.20$]$ & 2.00$\times10^{-2}$ & -8.16$\times10^{-3}$, +8.44$\times10^{-3}$ \\
$[$0.20, 0.25$]$ & 1.13$\times10^{-2}$ & -5.12$\times10^{-3}$, +5.06$\times10^{-3}$ \\
$[$0.25, 0.30$]$ & 3.83$\times10^{-3}$ & -2.61$\times10^{-3}$, +1.85$\times10^{-3}$ \\
$[$0.30, 0.40$]$ & 5.56$\times10^{-4}$ & -5.35$\times10^{-4}$, +3.86$\times10^{-4}$ \\
    \hline
    \hline
    \end{tabular*}
    \end{center}
\end{table*}

\begin{table*}[htbp]
    \begin{center}
    \caption{Production rate for the $\pizero$ production in the rapidity
    range $-10.0 > \ylab > -11.0$. \label{table:spectra_5}}
    \begin{tabular*}{14cm}{@{\extracolsep{\fill}}c c c}
    \hline
    \hline
    $\pt$ range $[$GeV$]$
    & Production rate $[$GeV$^{-2}]$
    & Syst+Stat uncertainty $[$GeV$^{-2}]$ \\
    \hline
$[$0.00, 0.05$]$ & -6.79$\times10^{-3}$ & -7.16$\times10^{-3}$, +7.43$\times10^{-3}$ \\
$[$0.05, 0.10$]$ & 6.12$\times10^{-3}$ & -4.74$\times10^{-3}$, +3.51$\times10^{-3}$ \\
$[$0.10, 0.15$]$ & 3.51$\times10^{-3}$ & -2.66$\times10^{-3}$, +1.54$\times10^{-3}$ \\
$[$0.15, 0.20$]$ & 2.01$\times10^{-3}$ & -1.39$\times10^{-3}$, +8.46$\times10^{-4}$ \\
$[$0.20, 0.30$]$ & 2.36$\times10^{-4}$ & -2.00$\times10^{-4}$, +1.08$\times10^{-4}$ \\
    \hline
    \hline
    \end{tabular*}
    \end{center}
\end{table*}

\section*{Appendix B}\label{sec:appendix_b}

\subsection*{Derivation of the $\pt$ spectra in $p$--$p$
collisions at $\sqs = \tev{5.02}$}\label{sec:pp_5TeV}

To investigate the nuclear effects involved in the nuclear target it is
essential to compare the $\pt$ spectra measured in $p$--\text{Pb} collisions at
a given collision energy to the reference $\pt$ spectra in $p$--$p$ collisions
at the same collision energy.
In this analysis, since a measurement in $p$--$p$ collisions at $\sqs =
\tev{5.02}$ is not available, the reference $\pt$ spectra are made by scaling
the $\pt$ spectra measured in the $p$--$p$ collisions at $\sqs = \tev{7}$ and
\tev{2.76}.

\begin{figure}[htbp]
  \centering
  \includegraphics[width=7cm, keepaspectratio]{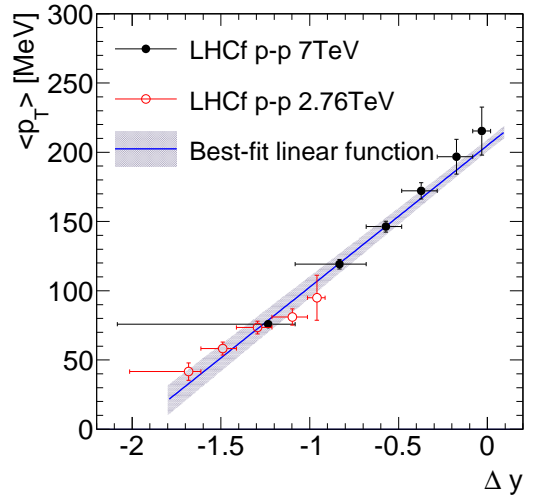}
  \caption{(color online). Average $\pt$ as a function of rapidity loss $\Delta
  y$. Filled circles indicate the LHCf results in $p$--$p$ collisions at $\sqs =
  $\tev{7} taken from Ref.~\cite{LHCfpi0}. Open circles indicate the LHCf
  results in $p$--$p$ collisions at $\sqs = $\tev{2.76}. The best-fit linear
  function to the LHCf data is shown by the solid line.}
  \label{fig:pt_scale}
\end{figure}

First the average $\pt$ at \tev{5.02} is estimated by scaling the average $\pt$
obtained at \tev{7} and \tev{2.76}. Figure~\ref{fig:pt_scale} shows the average
$\pt$ in $p$--$p$ collisions at $\sqs = \tev{7}$ (filled circles) and \tev{2.76}
(open circles) as a function of rapidity loss $\Delta y \equiv \ybeam - \ycms$,
where $\ybeam$ is beam rapidity and $\ycms$ is the rapidity of the
center-of-mass frame. For the proton beam with $E = \tev{3.5}$ and \tev{1.38},
$\ybeam$ gives 8.917 and 7.987, respectively. In the following we assume $\ycms$
is positive.

According to the scaling law proposed by several authors~\cite{Amati, Benecke,
Feynman} (Feynman scaling), the average $\pt$ as a function of $\Delta y$ should
be independent of the center-of-mass energy in the projectile fragmentation
region. Thus the average $\pt$ can be directly compared among different
collision energies. The values of the average $\pt$ at \tev{7} are taken from
measurements by LHCf~\cite{LHCfpi0} in which the associated $\Delta y$ points
are modified to take into account event population for each rapidity bin.
These weighted bin centers are estimated using the MC simulation by
\textsc{epos} 1.99. The values of the average $\pt$ at \tev{2.76} are obtained
by a similar analysis on the data that was taken in $p$--$p$ collisions at $\sqs
= \tev{2.76}$ on February 13, 2013. These data were taken with essentially the
same data acquisition configuration as at $\tev{5.02}$.

Although the two measurements in Fig.~\ref{fig:pt_scale} have limited overlap on
the $\Delta y$ range owing to the smaller collision energy at \tev{2.76}, the
$\langle\pt\rangle$ spectra at $\tev{7}$ and $\tev{2.76}$ follow mostly a common
line. A linear function fit is then made to these measurements.
The solid line and shaded area in Fig.~\ref{fig:pt_scale} show the best-fit
linear function and the 1 standard deviation uncertainty obtained by a
chi-square fit to the data points
\begin{equation}
 	\langle \pt \rangle_\text{best-fit}(\Delta y) = 216.3 + 116.0\Delta y.
 	\label{eq:ave_pt}
\end{equation}
\noindent The minimum chi-square value is 11.1 with a number of degrees of
freedom equal to 9.
With the fitted result in Eq.~\ref{eq:ave_pt} and $\ybeam$ for the proton beam
with $E = \tev{2.51}$, 8.585, the average $\pt$ at a given rapidity $\ycms$ at
\tev{5.02} can be evaluated as
\begin{equation}
 	\langle \pt \rangle(\ycms)|_{\tev{5.02}} = 216.3 + 116.0(8.585 - \ycms),
 	\label{eq:ave_pt_5TeV}
\end{equation}
\noindent where we assume the proton beam travels to the positive rapidity
direction. Note that the rapidity range of the reference $\pt$ spectra at
\tev{5.02} is enclosed by the data points taken at \tev{7} and \tev{2.76}.

The absolute normalization scaling among three collisions is then estimated for
\tev{2.76}, \tev{5.02}, and \tev{7} energies. Since the systematic uncertainty
of the LHCf measurements on the luminosity is $\pm\SI{20}{\percent}$ and
$\pm\SI{6.1}{\percent}$ at $\tev{2.76}$ and $\tev{7}$ respectively, the
predictions by MC simulations are used instead for the estimation.
According to \textsc{dpmjet} 3.04, {\sc qgsjet} II-03 and \textsc{epos} 1.99,
the relative normalization at \tev{5.02} to at \tev{7} or \tev{2.76}, defined as
\begin{align}
	R_\text{norm} \equiv
	&\left.\int dp^3\frac{1}{\sigma_\text{inel}}
	E\frac{d^3\sigma}{dp^3}\right|_{\sqs=\tev{5.02}} \\ \nonumber
	&\left/\int dp^3 \frac{1}{\sigma_\text{inel}}
	E\frac{d^3\sigma}{dp^3}\right|_{\sqs=\tev{7}~\text{or}~\tev{2.76}},
\end{align}
is mostly unity in the rapidity and $\pt$ ranges covered by LHCf. Therefore we
apply the measured absolute normalization at \tev{7} to the reference $\pt$
spectra at \tev{5.02} without scaling. The uncertainty on the normalization is
taken from the luminosity error $\pm\SI{6.1}{\percent}$.

Accordingly the average $\pt$ and normalization of $\pt$ spectra at \tev{5.02}
can be scaled from \tev{7} and \tev{2.76}. With these two values, the $\pt$
spectra in $p$--$p$ collisions at \tev{5.02} can be effectively derived. In the
analysis of this paper, the expected $\pt$ spectra are presumed to follow a
Gaussian distribution with the width of the distribution $\sigma_\text{Gauss}$
equal to $2\langle \pt \rangle/\sqrt{\pi}$. The expected $\pt$ spectra in
$p$--$p$ collisions take into account the rapidity shift $-0.465$ explained in
Sec.~\ref{sec:data} for consistently with the asymmetric beam energies in
$p$--Pb collisions.

\end{document}